\documentclass[a4paper,11pt]{article}
\pdfoutput=1 

\usepackage{jinstpub} 
\usepackage{comment}

\title{Advanced Assessment of Beam-Induced Background at a Muon Collider}

\author[a]{Francesco Collamati,}
\author[b,f,1]{Camilla Curatolo,\note{Corresponding author.}}
\author[b]{Donatella Lucchesi,}
\author[c]{Alessio Mereghetti,}
\author[d]{Nikolai Mokhov,}
\author[e]{Mark Palmer,}
\author[f]{Paola Sala}

\affiliation[a]{INFN Sezione di Roma, Roma, Italy}
\affiliation[b]{University of Padova and INFN Sezione di Padova, Padova, Italy}
\affiliation[c]{CERN, Geneva, Switzerland, currently at Fondazione CNAO, Pavia, Italy}
\affiliation[d]{Fermilab, Batavia, Illinois, United States}
\affiliation[e]{Brookhaven National Laboratory, Upton, New York, United States}
\affiliation[f]{INFN Sezione di Milano, Milano, Italy}

\emailAdd{camilla.curatolo@pd.infn.it}

\abstract{Renewed international interest in muon colliders motivates the continued investigation of the impacts of beam-induced background on detector performance.  This continues the effort initiated  by the Muon Accelerator Program and carried out until 2017. The beam-induced background from muon decays directly impacts detector performance and must be  mitigated by optimizing the overall machine design, with particular attention paid to the machine detector interface region. 
In order to produce beam-induced background events and to study their characteristics in coordination with the collider optimization, a flexible simulation approach is needed. To achieve this goal we have chosen to utilize the combination of LineBuilder and Monte Carlo FLUKA codes. We report the results of beam-induced background studies with these tools obtained for a 1.5 TeV center of mass energy collider configuration. Good agreement with previous simulations using the MARS15 code demonstrate that our choice of tools meet the accuracy and performance requirements to perform future optimization studies on muon collider designs.
}

\keywords{Accelerator Subsystems and Technologies. Detector modelling and simulations I (interaction of radiation with matter, interaction of photons with matter, interaction of hadrons with matter, etc).}

\arxivnumber{2105.09116}

\begin{document}
\maketitle
\flushbottom

\section{Introduction}

In the ongoing endeavor to deepen our understanding of the fundamental structure of matter, muon colliders (MCs) represent an alternative approach to classical stable particle ($pp$ and $e^+e^-$) colliders. Although the LHC still has several years of data to deliver, work is underway now to explore new machine options to carry on the research on fundamental particle physics in the next decades.
Despite having been first proposed in the late `60s~\cite{ref:mucoll_history1,ref:mucoll_history2}, and studied in detail by the Muon Accelerator Program (MAP)~\cite{ref:MAP}, the MC concept is receiving renewed interest from the scientific community~\cite{cm1, cm2} because of its potential to overcome key limitations of other proposed collider concepts. 

With respect to the next generation of electron-positron colliders being considered, machines that aim to deliver unprecedented lepton collision energies and luminosities (e.g., the Future Circular Collider (FCC-ee)~\cite{ref:FCC}, International Linear Collider (ILC)~\cite{ref:ILC} and Compact Linear Collider~\cite{ref:CLIC} design studies), colliding muons can deliver key advantages for accessing the energy frontier of high energy physics. With the muon being about 200 times heavier than an electron, synchrotron radiation is expected to be almost negligible for muon  beams up to very high energies, thus allowing a circular muon collider to reach the multi-TeV regime.  In contrast, synchrotron radiation places a fundamental limit on the energy reach of circular $e^+e^-$ colliders. Linear $e^+e^-$ colliders suffer from beamstrahlung \cite{ref:beamstr}, which reduces the actual luminosity and generates unwanted beam--$\gamma$ interactions. Colliding muons enables a reduced energy spread in the collisions resulting in significantly improved energy resolution for physics measurements.

Work carried out so far has not fully investigated the potential physics reach of proposed MC designs. A first comparison of possible measurements with respect to other future colliders can be found in Ref.~\cite{ref:theory_SG}, which focuses  on a machine with a 10~TeV center of mass (CM) energy, an energy where new physics could be tested beyond the limits of any other future collider. An evaluation of the physics potential of a MC at 3~TeV CM energy is still in progress, but preliminary results~\cite{ref:Nazar2} show its competitiveness with more mature projects such as CLIC. 

The tremendous physics potential of colliding muon beams is tempered by the challenges of constructing a collider using unstable particles.  The short lifetime of the muon requires efficient and rapid production, manipulation and acceleration of the muon beams.  In particular, tertiary production of muons from protons striking a target, as studied by the MAP collaboration, requires significant cooling of the beam emittance via the ``ionization cooling'' process that has been recently characterized by the Muon Ionization Cooling Experiment (MICE)~\cite{ref:mice}. An alternative approach to produce low emittance muon beams~\cite{ref:lemma} has been proposed by the LEMMA collaboration, but requires considerable additional study.

Challenges in the MC complex include the operation of the machine and the detector in an environment where the beams rapidly decay. In particular, muon decays all along the collider result in a continuous flux of secondary and tertiary particles, the so--called ``Beam--Induced Background'' (BIB), that could jeopardize the physics performance of the detector if not properly dealt with. 

The amount of BIB impacting the detector depends on the CM energy and on the instantaneous luminosity, key figures of the IR design. Each machine configuration requires a dedicated optimization of the MDI region, which requires a seamless transition between machine optics and Monte Carlo simulations. As demonstrated by the first comprehensive BIB studies~\cite{ref:mok2012} performed with MARS15, the MDI optimization is expected to be an iterative process, where changes in the machine optics, even hundreds of meters away from the IP, can substantially affect the BIB distribution in the detector.

In this paper a flexible and advanced approach to BIB simulation is presented, based on the FLUKA Monte Carlo code~\cite{ref:FLUKA1, ref:FLUKA2}; results for the 1.5~TeV CM energy case based on the MAP design of the collider are shown. A quantitative benchmark against MARS15 results for the same configuration is reported, showing the reliability of the simulation tool.

\section{Relevance of the Beam--Induced Background}

BIB can have detrimental impacts on several elements of the accelerator complex, as discussed in Ref.~\cite{ref:mok2012_2}. It also affects the detector performance due to the significant flux of muon decay products, with a broad energy spectrum, that are generated throughout the ring.  In the absence of adequate shielding, the tracking system, which is the detector portion closest to the interaction point (IP), would suffer high occupancy due to the charged secondary and tertiary muon decay products. The inner layers would be not usable, thus making it impossible to perform the high precision track reconstruction that is required for the identification of secondary vertices of short--lived particles like the b--hadrons. The huge flux of neutrons  striking the calorimeter detector would generate large energy deposition, similar to a high density underlying event, thus degrading the jet and electron/positron identification performance beyond that needed for high quality physics measurements.

The first detailed design and optimization of the Interaction Region (IR) and Machine--Detector Interface (MDI) are described in Refs.~\cite{ref:Alex, ref:mok2012_2}. The key element of that design, the deployment of nozzles as shielding elements, was originally proposed in Ref.~\cite{ref:foster}. The nozzles are double--cone--shaped tungsten absorbers, located inside the detector in the immediate vicinity of the IP. These absorbers reduce the BIB in the detector by orders of magnitude~\cite{ref:mok2012_2, ref:mok2012}, resulting in a detector performance in line with other future collider experiments. 

The effect of the nozzles has been demonstrated by means of simulations performed with the MARS15 Monte Carlo code~\cite{ref:MARS15, ref:mok2012} in the context of the MAP program. The dimensions and shape of the shields have been optimised taking into account that additional mitigation strategies are possible at the level of detector and IR design~\cite{ref:mok2012,ref:diben}.

\section{Simulation of 1.5 TeV BIB}
\subsection{Simulation Setup}

The source of BIB is the decay of primary muons and the interaction of the decay products with collider and detector components. Radiation fields of this type are commonly simulated by Monte Carlo codes, one of the most used being FLUKA. FLUKA is a general purpose Monte Carlo code, used in a variety of applications ranging from medical to astro--particle physics. In particular, FLUKA is the golden standard in radioprotection and shielding studies.

FLUKA natively supports very complicated geometries, thus allowing  accurate modeling of the full accelerator complex in the simulation. However, the manual construction of such complex geometries can be both difficult and error prone, and, above all, it is definitely not compatible with the flexibility required by this particular application.

The possibility to model in FLUKA the accelerator lattice and optics optimized with codes like MAD--X~\cite{ref:madx} is offered by the Python code  FLUKA LineBuilder (LB)~\cite{ref:LineBuilder}. The LB automatically constructs the FLUKA geometry of the accelerator beam line according to the optics, placing each needed accelerator element at the proper position with the correct orientation and magnetic fields.

In this work, the 1.5~TeV CM energy machine configuration is used as a case study to validate these tools since an existing IR configuration optimized by MAP along with a BIB sample generated with MARS15 are available for quantitative comparison.

\subsection{Implementation}
\begin{figure}[htbp]\centering
	\includegraphics[width=1.\textwidth]{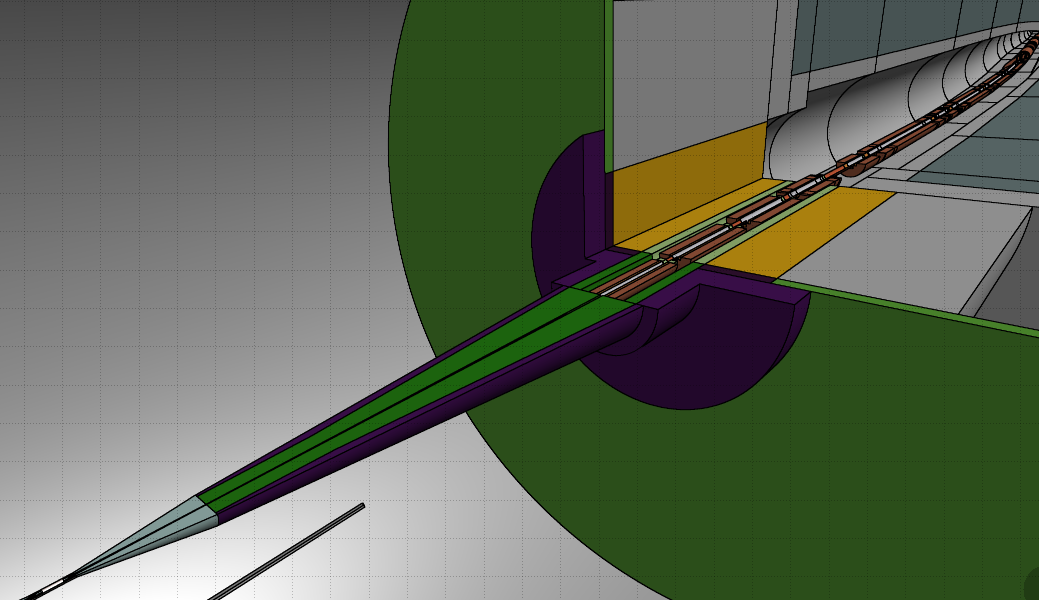}
	\caption{\label{3Dgeo} 3D visualization (horizontal cut at beam height) by means of FLAIR~\cite{ref:FLAIR} of the FLUKA geometry of the MC IR simulated in the present work.}
\end{figure}
Figure~\ref{3Dgeo} shows a 3D rendering of the FLUKA geometry assembled with the LB; the IR with the nozzle and the first magnets around the IP are shown.
The IP is located at the origin of the global reference system of FLUKA, with the x--axis pointing outside the ring, the z--axis tangent to the ring and the y--axis pointing vertically up; the $\mu^-$ beam travels through the accelerator ring counterclockwise, i.e.~opposite the z--axis at the IP.
Families of accelerator elements have been defined in a database according to the information contained in the MAD8~\cite{MAD8_user_manual,MAD8web} optics file and to those reported in MAP publications, in particular in Ref.~\cite{ref:Alex}. Residual discrepancies in the geometry and materials with respect to MAP studies are possible. The parameter list of the magnetic elements close to the IP are reported in Table~\ref{mags}.
\begin{table}[htbp]
	\centering	
	\caption{\label{mags} Parameters describing the magnets close to the interaction point.  $Q_1$, $Q_2$, $Q_3$, $Q_4$, $Q_5$ are quadrupoles, whereas $B_1$ is a dipole. The coils are delimited by the inner and outer radius, R$c_{int}$ and R$c_{out}$ respectively, and the whole magnet is delimited by the outer radius of the iron yoke R$i_{out}$. }
	\smallskip
	\begin{tabular}{cccccc}
	\hline
	\textbf{Name}   & \textbf{$Q_1$} & \textbf{$Q_2$}  & \textbf{$Q_3$, $Q_4$} & \textbf{$Q_5$} & \textbf{$B_1$}\\
	\hline
	R$c_{int}$ (cm) & $4$ & $5.5$ & $8$ &$8$ &$8$ \\
	R$c_{out}$ (cm)& $8$& $9.5$ & $12$ &$12$&$12$\\
	R$i_{out}$ (cm)  & $20$ & $25$ & $30$ & $30$& $30$ \\
	Length (m) & $1.5$  &  $1.76$  & $1.7$  &$1$ &$6$ \\
	\hline
	\end{tabular}		
\end{table}

\begin{figure}[htbp]\centering
	\includegraphics[width=1.\textwidth]{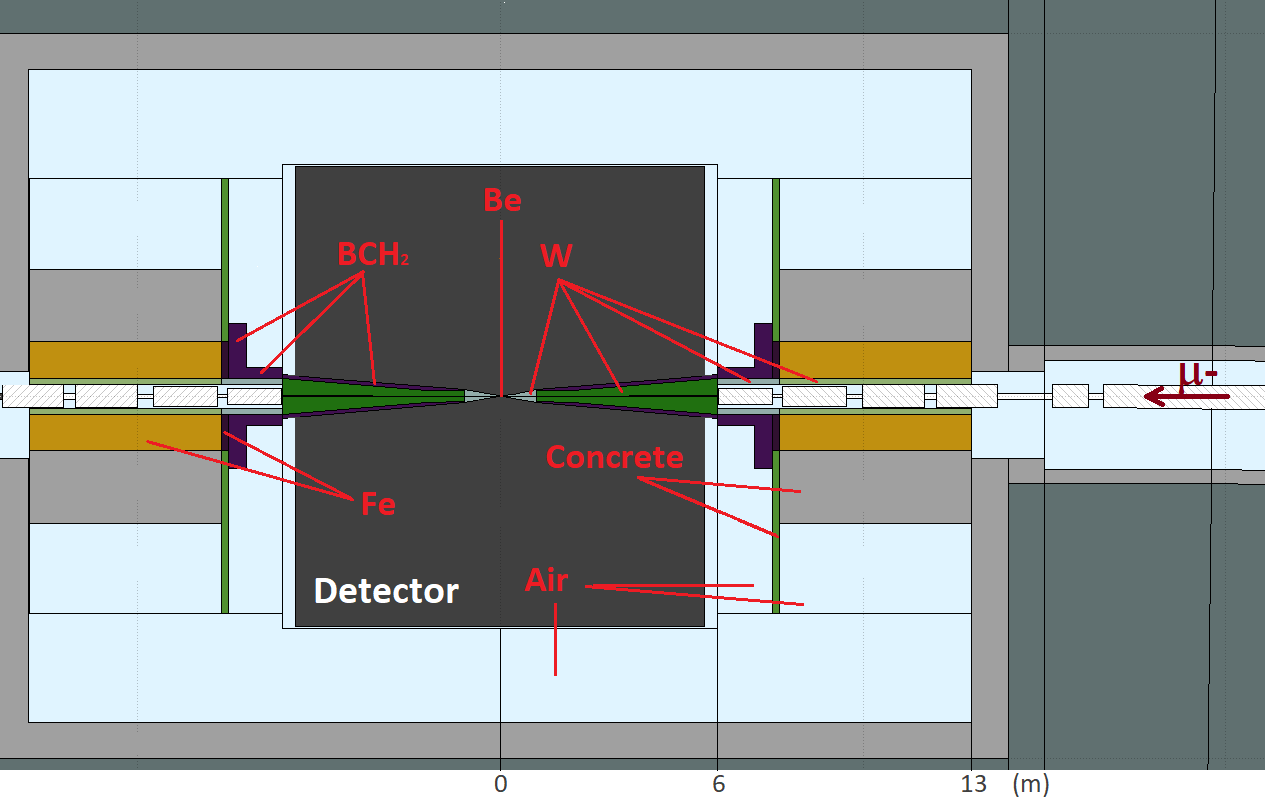}
	\caption{\label{ir}Interaction region. The passive elements, the nozzles and the pipe around the interaction point are constituted by iron (Fe), borated polyethylene (BCH$_2$), berillium (Be), tungsten (W) and concrete. The detector outer shape is a $11.28$~m long cylinder of $6.3$ m radius. The space between the outer shape and the nozzles is considered as a perfect particle absorber (``blackhole''). The bunker is a $26$~m-long cylinder with a radius of $9$~m.}
\end{figure}

\begin{figure}[htbp]\centering
	\includegraphics[scale=0.7]{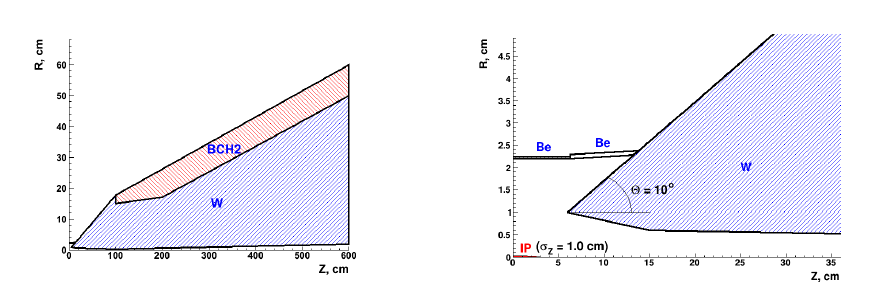}
	\caption{\label{nozzle} Detailed geometry and material description of the nozzle from \cite{ref:diben}.}
\end{figure}

The representation of the IR and the material of the components are displayed in Fig.~\ref{ir}, while the nozzle geometry is shown in Fig.~\ref{nozzle}. A solenoidal magnetic field of 3.57~T is present in the IR.

The relevant quantity of interest for this study is the flux of particles exiting from the machine; the detailed study of what happens inside the detector is then performed by a separate simulation~\cite{ref:Nazar2} and is not part of the presented study. 

Given the symmetric nature of the $\mu^+$/$\mu^-$ collider,  only one primary beam is considered. In particular, the primary $\mu^-$ beam is simulated travelling counterclockwise starting 200~m away from the IP (in the pictures the primary beam arrives from the right, see for example Fig.~\ref{ir}). The sampled muon beam is monochromatic, at 750~GeV/c.
On both transverse planes, it has a Gaussian distribution both in position and direction, with standard deviations equal to 5.96~$\mu$~m and 596~$\mu$rad, respectively; these values reflect the beam optics functions at the interaction point, i.e.~$\beta$=1~cm and $\alpha$=0, for a normalised transverse emittance of 25~$\pi$~mm~mrad corresponding to a geometric transverse emittance of 3.5~$\pi$~nm~rad~\cite{ref:Alex}.
The bunch intensity is assumed to be $2\times10^{12}$.
These values are reported in Table~\ref{tab:beamRef}.
\begin{table}[h]
	\centering
    \caption{\label{tab:beamRef}Key figures of the machine configuration and beam properties used in the BIB estimation.}
    \smallskip
    \begin{tabular}{lrl}
		\hline
		\textbf{Parameter}  & \textbf{Value} & \\ 
		\hline
		Beam momentum & 750 & GeV/c \\
		Beam momentum spread & 0 & GeV/c \\
		Bunch intensity & $2 \times10^{12}$ &  \\
		$\epsilon^n_{x,y}$ normalised RMS emittance & 25 $\pi$  $10^{-6}$  & m rad \\
		$\epsilon^g_{x,y}$ geometric RMS emittance & 3.5 $\pi$  $10^{-9}$   & m rad \\
		\hline
		$\beta_{x,y}$ & 1 & cm \\
		$\alpha_{x,y}$ & 0 & \\
		\hline
		$\sigma_{x,y}$ RMS beam size& 5.96 & $\mu$m \\
		$\sigma_{x',y'}$ RMS beam divergence & 596 & $\mu$rad \\
        \hline
	\end{tabular}	
\end{table}

In order to have a reasonable statistics, the muon decay has been biased, increasing the decay probability; decay products are assigned a weight later used for estimating results, to compensate for the bias. Decay products are further transported in the geometry, with accurate description of electromagnetic and hadronic processes. Hadrons (mostly neutrons) are generated through electronuclear and photonuclear interactions.

The scoring is performed by saving in a dump file tracks exiting the machine, either from the tungsten nozzle or from any of the IR components. In particular, these quantities are registered: \textit{particle type, energy, momentum, statistical weight, position, time, position of the decaying muon, region of the machine where the first interaction of secondary particles occurs and machine region from where the particle exits}. Energy, momentum and direction are given at the machine exit surface. Time is defined with respect to the interaction time at the IP.

A Python software algorithm has been developed to analyze the simulation output and to study the BIB properties before the interaction with the detector. This is an important element in view of the flexibility imposed by the studied case, since the iterative task of MDI optimization will rely on a prompt interpretation of the Monte Carlo simulation output, for example when the effect of a single modification in the lattice must be evaluated.

\begin{figure}[h]
\centering
\includegraphics[width=1.\textwidth]{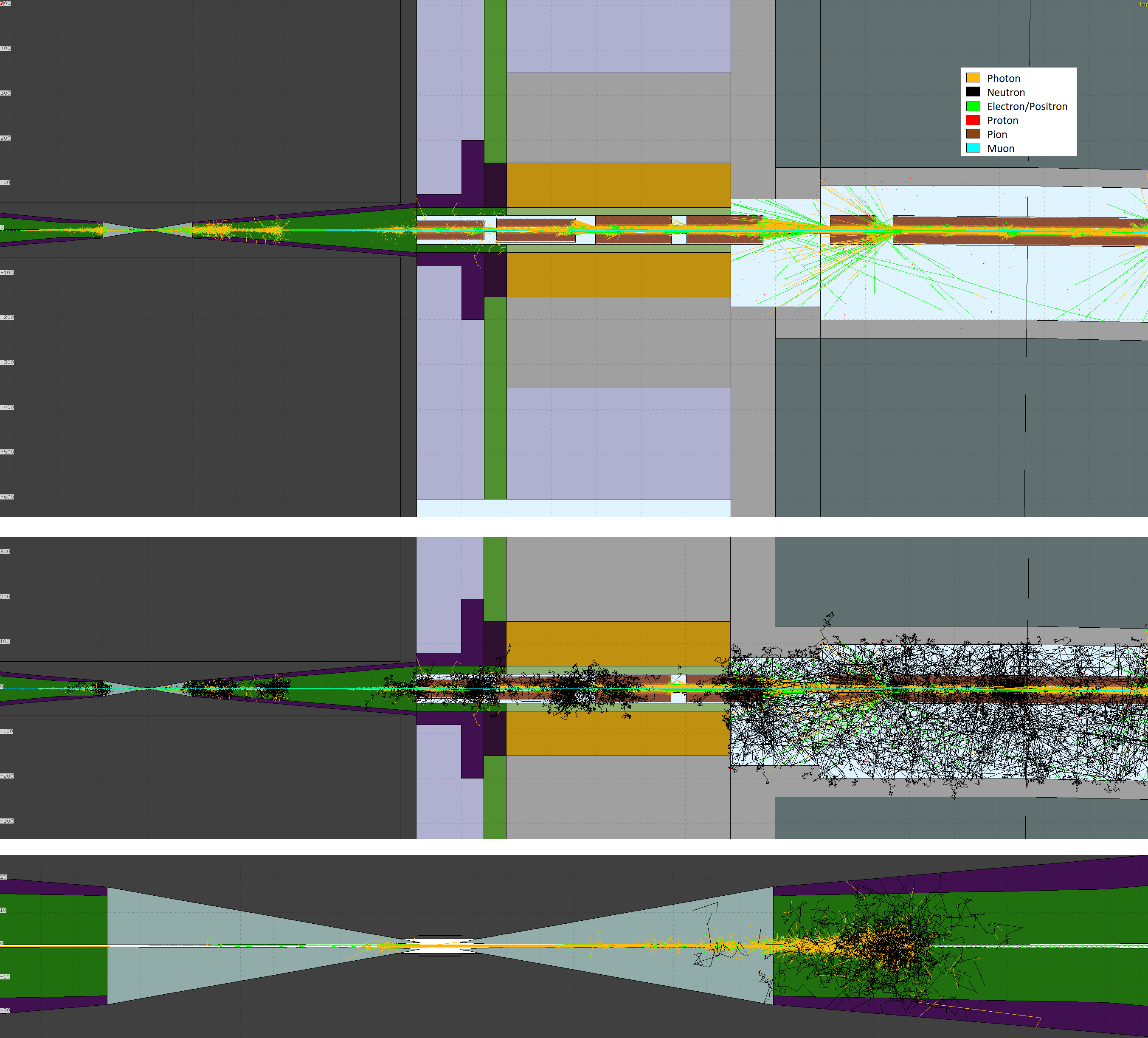}
\caption{\label{tracks} Pictorial view of tracks of secondary particles in the IR and in the first magnets around the IR in case of few muon decays: all particles but neutrons (top frame) and all particles including neutrons (middle frame). Bottom frame: secondary particle tracks in case of a single muon decay in the proximity of the IP.}
\end{figure}
A pictorial view of the complexity of the simulated events in the last tens of meters from the IP is shown in Fig.~\ref{tracks}. The top picture shows examples of tracks of secondary particles except neutrons for a few muon decays, while in the middle one neutrons are included. The bottom plot illustrates the tracks in the case of a single muon decay in the proximity of the IP.

\section{Comparison between FLUKA and MARS15 results}
In this section, the total number of particles and the energy, time and space distributions obtained with FLUKA are compared to those obtained by MARS15 and mostly reported in \cite{ref:mok2012, ref:mok2012_2, ref:diben}.
The comparison considers only primary muons decaying within 25~m from the IP as in the available MARS15 dataset. The energy threshold cuts are set as stated in Ref.~\cite{ref:diben}, reported here for the redear's convenience: 100~keV for $\gamma$, $e^+/e^-$, charged hadron (proton, $\pi^+/\pi^-$, $K^+/K^-$), $\mu^+/\mu^-$ and 1~meV for neutron.

\begin{figure}[h]\centering
\includegraphics[width=1.\textwidth]{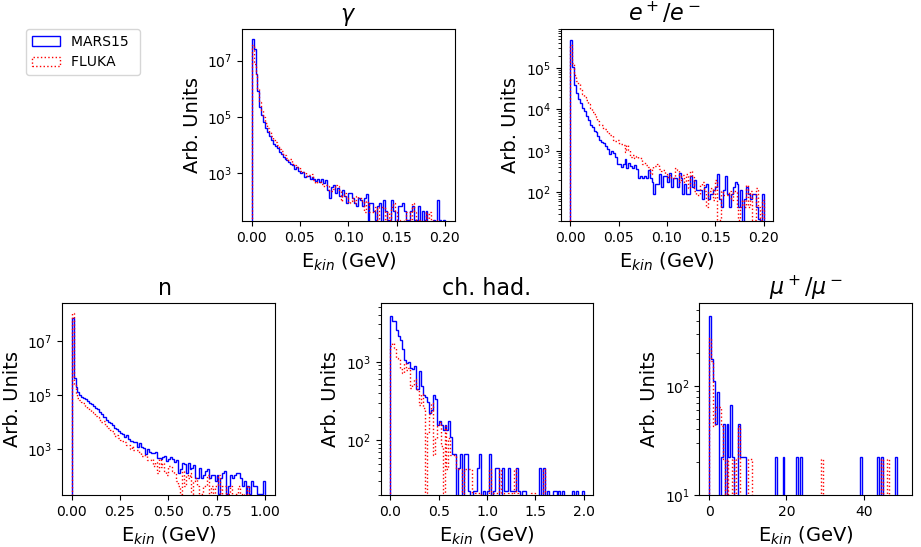}
	\caption{\label{energy}Energy distribution of BIB particles obtained with MARS15 in solid blue line and FLUKA in dotted red line. Results for a $2\times10^{12}$~$\mu^-$ beam, decaying within 25~m from the IP.}
\end{figure}

\begin{figure}[h]\centering
	\includegraphics[width=1.\textwidth]{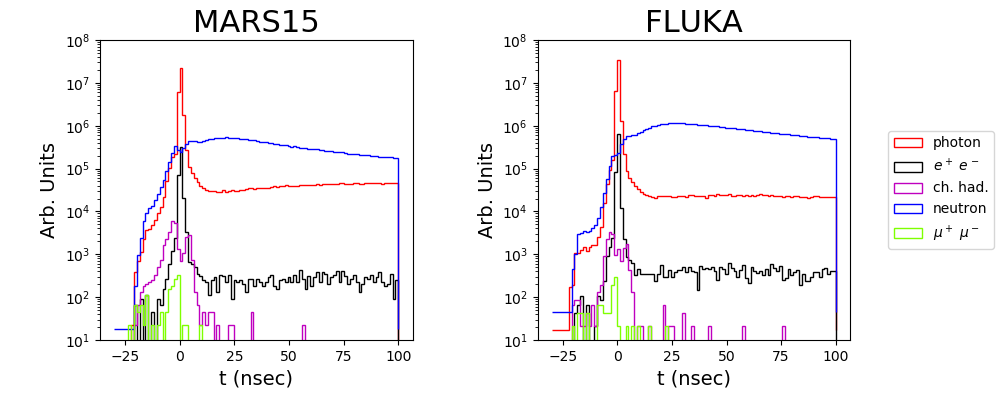}
	\caption{\label{time}Time distribution up to 100 nsec of BIB particles simulated with  MARS15 (left plot) and FLUKA (right plot). Results for a $2\times10^{12}$~$\mu^-$ beam, decaying within 25~m from the IP.}
\end{figure}

\begin{figure}[h]\centering 
	\includegraphics[width=1.\textwidth]{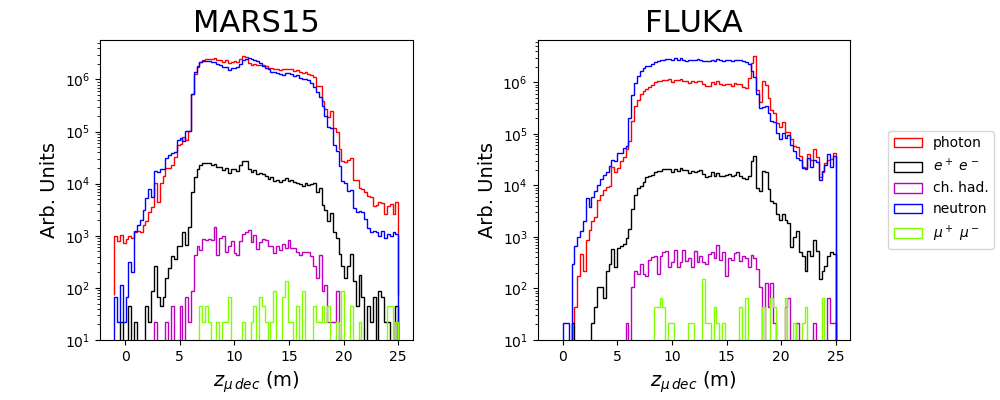}
	\caption{\label{mudec}Distribution of particle type entering the detector as a function of $z_{\mu}$, the longitudinal coordinate of the primary muon decay. Results for a $2\times10^{12}$~$\mu^-$ beam, decaying within 25~m from the IP.}
\end{figure}

Table~\ref{tab:num} reports the breakdown by particle type of the total number of particles that exit the machine produced by the primary muons' decay within 25~m from the IP; the corresponding energy and time distributions are shown in Figs.~\ref{energy} and~\ref{time}. The distributions of the particles exiting the machine as a function of longitudinal primary muon decay coordinate are reported in Fig.~\ref{mudec}. All results are normalized to a beam intensity of 2~$\times10^{12}$~muons.

\begin{table}[h]
	\centering
    \caption{\label{tab:num} Number of BIB particles obtained using MARS15 and FLUKA. For each particle type the threshold energy is also reported. Results for a $2\times10^{12}$~$\mu^-$ beam, decaying within 25~m from the IP.}
    \smallskip
    \begin{tabular}{ccc}
		\hline
		\textbf{Particle ($E_{th}$)}& \textbf{MARS15} & \textbf{FLUKA} \\ 
		\hline
		Photon ($100$ keV)            & $8.6 \, 10^7$  & $5 \, 10^7$ \\ 
		Neutron (1 meV)           &	$7.6 \, 10^7$  & $1.1 \, 10^8$   \\
		Electron/positron ($100$ keV) & $7.5 \, 10^5$  & $8.5 \, 10^5$  \\ 
		Ch. Hadron ($100$ keV)          & $3.1 \, 10^4$  & $1.7 \, 10^4$   \\
		Muon ($100$ keV)                &	$1.5 \, 10^3$  & $1 \, 10^3$  \\
		\hline
	\end{tabular}	
\end{table}

The major contributors to the BIB are photons, neutrons and electrons/positrons, as confirmed by both simulation programs. The kinetic energy and time distributions of all particles have similar shapes. Overall, the number of particles predicted by FLUKA and MARS15 agree within a factor of two or better.
It should be kept in mind that the two simulations do not use exactly the same IR geometry, material composition and configuration, due to the difficulties in retrieving in detail such information.
As a matter of fact, a previous comparison~\cite{ref:MvsF125} performed by the MAP collaboration between MARS15 and FLUKA results of the BIB generated at $125$~GeV CM energy had highlighted a remarkable agreement between the two codes when the two geometries are modelled with the same level of accuracy.

\begin{figure}[h]\centering
	\includegraphics[width=1.\textwidth]{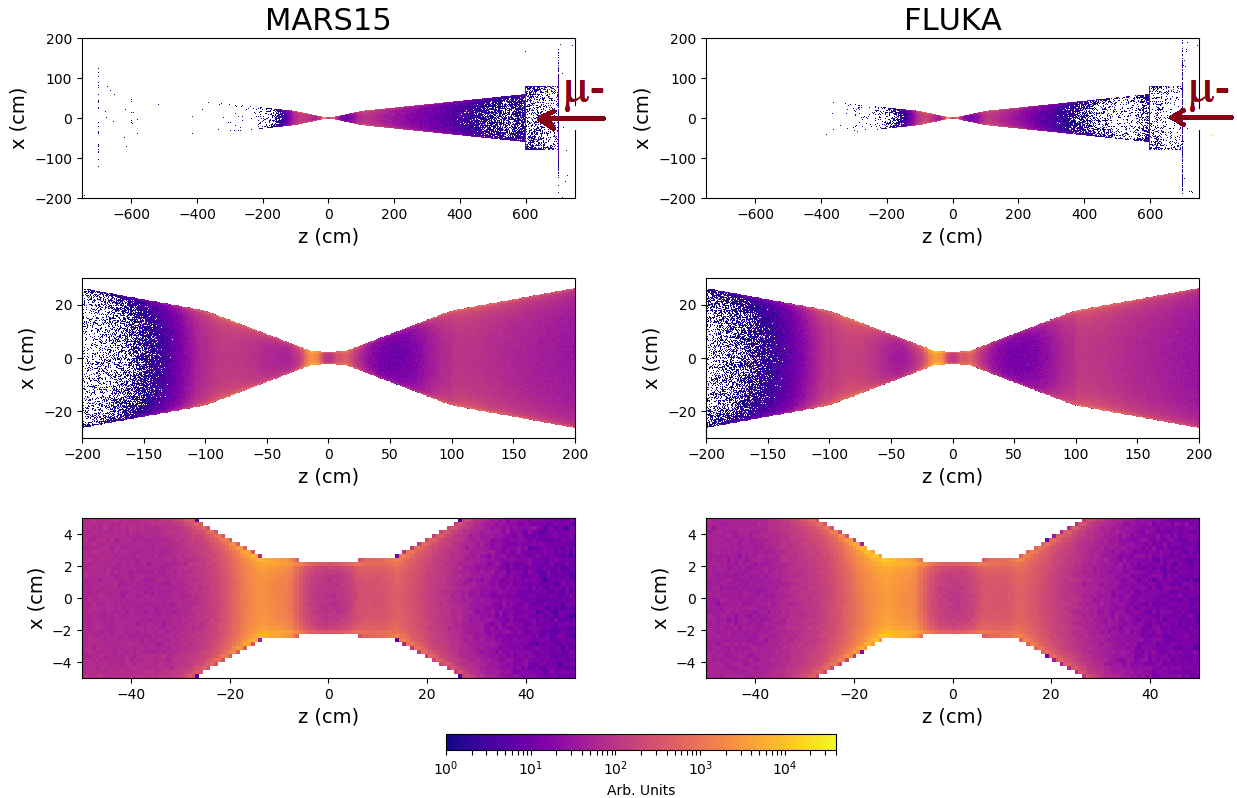}
\caption{\label{colormap}Colormap of the particles entering the detector represented in the x,z plane. From top to bottom a zoom in around the IP is shown. Results for a $2\times10^{12}$~$\mu^-$ beam, decaying within 25~m from the IP.}
\end{figure}

Fig.~\ref{colormap} shows a color--map of the exit point of the particles in the x--z plane, for the $\mu^-$ beam, coming from the right. Most of them exit from the beam pipe and the nozzles close to the IP, as is particularly visible in the magnified bottom row of plots. The beam reaching the nozzle from the right is collimated towards the left tip of the nozzle where a large number of interactions occurs. This evidence demonstrates the need to investigate the origin of the BIB in the IR elements, for how far in the machine geometry we should consider muon decays and what is the effect of the nozzles. These issues are discussed in detail in Sections~\ref{sec:origin} and~\ref{sec:noz}.

\section{Study of the BIB origin} \label{sec:origin}
\begin{figure}[h]\centering
	\includegraphics[width=1.\textwidth]{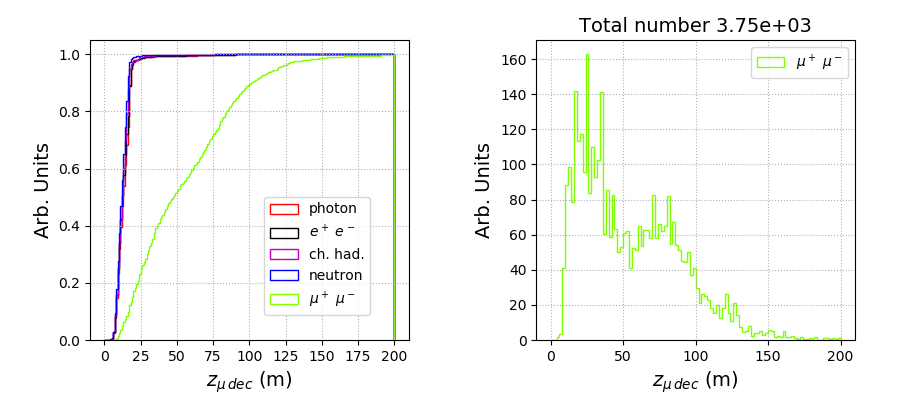}
	\caption{\label{mudec3} Cumulative distribution of particles exiting the machine by type, as a function of the primary muon decay longitudinal coordinate (left plot), and distribution of primary muon decays generating secondary muons (right plot). The total number of secondary muons is reported in the title.}
\end{figure}

Figure~\ref{mudec3} (left plot) shows the cumulative distributions of particles exiting the machine as a function of the primary muon decay longitudinal coordinate. As it can be seen, all particles exiting the machine towards the detector hall originate from muons that decay within $\approx25$~m from the IP. On the contrary, in order to have a correct estimate of the secondary muons at the IP, it is necessary to consider primary muons decay up to 150~m (right plot), as already pointed out by MARS15 studies~\cite{ref:mok2012}.

\begin{figure}[h]\centering
	\includegraphics[width=1.\textwidth]{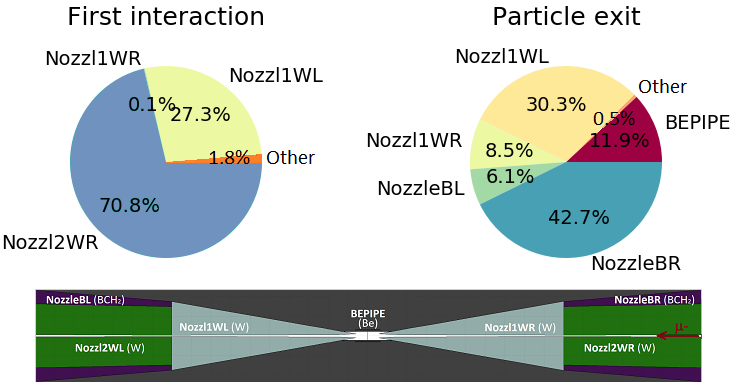}
    \caption{\label{regions} Top--left frame: elements of the IR where the first interactions occur after the primary muon decay. Top--right frame: elements from which the BIB particles exit the machine. Bottom frame: sketch of the IR with the relevant region names and materials. Results for a $2\times10^{12}$~$\mu^-$ beam, decaying within 25~m from the IP.}
\end{figure}

The major sources of BIB along the accelerator complex have been studied in detail. 
In particular, the top--left frame of Fig.~\ref{regions} shows the breakdown by region where the decay products undergo their first interaction. As expected, the majority occur in the right nozzle, but a non-negligible number also occur in the left nozzle. 
Moreover, the pie--chart in the top--right frame of the same figure shows the breakdown by region from which the BIB particles exit the machine. Among all the elements, the right nozzle is the most relevant. The sketch at the bottom displays the element names and materials. These three graphs together demonstrate that, noting that the primary beam arrives from the right, the first interactions occur mainly in the right part of the nozzle that, together with the right tip, acts as an absorber and collimator, and on the opposite nozzle tip, that operates as a target, from which $\sim30\%$ of BIB exits the machine.

\section{BIB characteristics without the nozzle absorber}\label{sec:noz}
\begin{figure}[h]
\centering
\includegraphics[width=1.\textwidth]{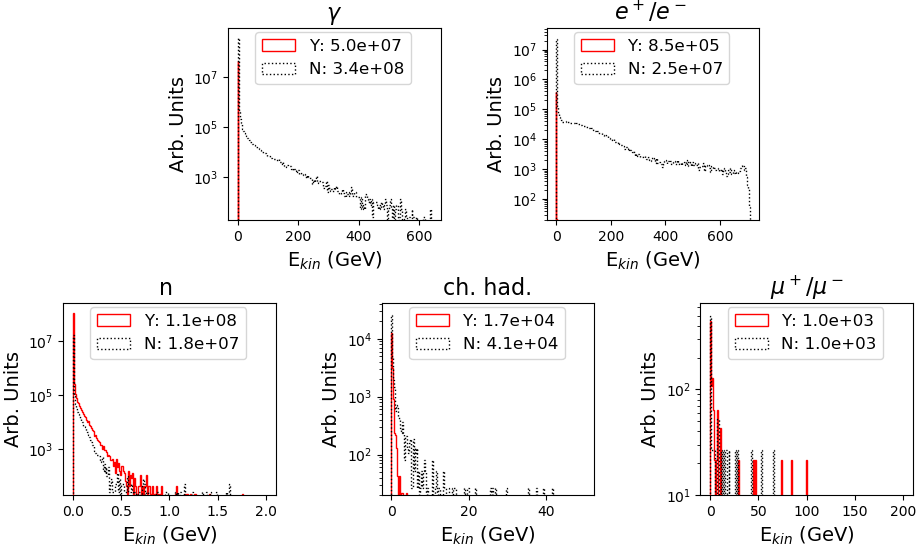}
\caption{\label{NoNozzle} Comparison of number and energy spectra of the BIB: with nozzles (Y) in solid red line and without nozzles (N) in dotted black line.}
\end{figure}

In order to better assess the role of the nozzles, a simulation of BIB without them has been carried out.
Figure~\ref{NoNozzle} shows the comparison between energy distributions and total fluxes "with nozzles" (Y) and "without nozzles" (N).
As expected, a major increase in particle flux with the nozzles removed is observed for photons and $e^+ e^-$, whereas a milder increase is observed for charged hadrons and muons flux; in contrast, the neutron flux decreases.
Along with the total number of particles, it is also important to highlight how the energy spectra are affected, which in absence of the nozzles reach very high energies. In particular, without the nozzles, the electromagnetic component would completely jeopardize the detector performance; hence, the nozzles play a crucial role in suppressing this particular component.

\section{Radiation levels in the detector}
The successful modeling of the radiation environment in the MDI allows reuse of the same setup for an evaluation of radiation hazards to the various detector components. 
With this goal, a simplified geometry of the detector, in agreement with Ref.~\cite{ICHEP2020-Casarsa}, has been implemented in FLUKA. All the silicon layers composing the inner tracker have been included with exact dimensions. The calorimeters have been approximated with cylindrical elements having  density and material composition matching the averages from the real ones (Si--W layers for the electromagnetic calorimeter, steel--scintillator for the hadronic one). The same approximation has been implemented for the magnetic coils and the return yoke. A uniform solenoidal field of $3.57$~T is present in the detector region, and a field of $1.34$~T circulates in the joke.

There exists various metrics defining radiation fields and hazard for detectors and associated electronics. The total ionizing dose is surely one of them. For silicon based detectors, where displacement of silicon atoms from their lattice position is an important source of malfunctioning, the radiation fields is customarily characterised in terms of the so--called 1~MeV neutron equivalent fluence (1-MeV-neq), meaning that the particle fields are converted to the fluence of 1~MeV neutrons that would produce the same damage.
FLUKA provides the capability to score by  online  convolution of particle fluences with conversion tables. 

\begin{figure}[h]
\centering
\includegraphics[width=1.\textwidth]{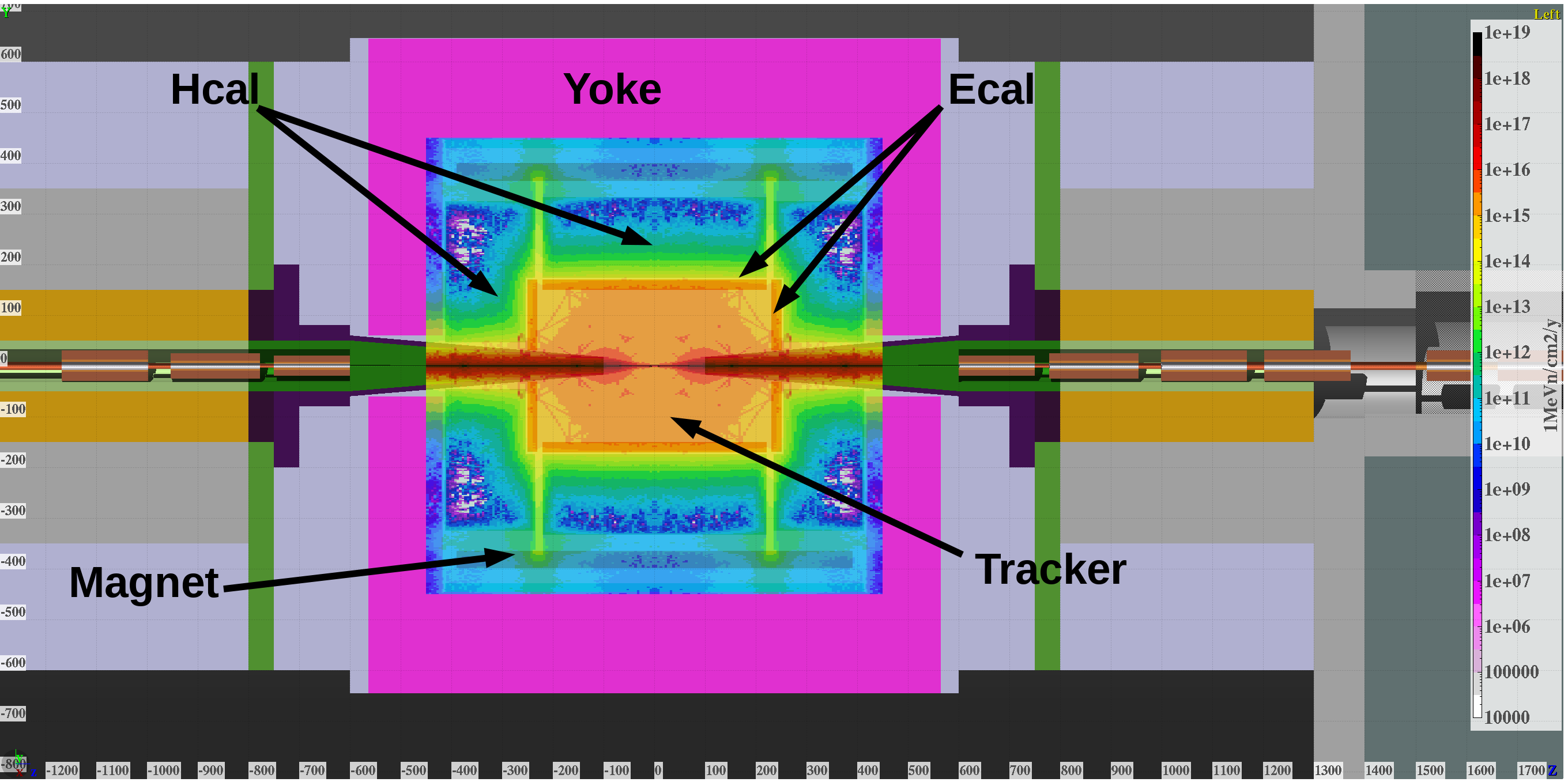}
\caption{\label{1MeVneq} 1~MeV neutron equivalent fluence in the detector region, normalized to one year of operation}
\end{figure}
The map of 1-MeV-neq in the region internal to the yoke is shown in Fig.~\ref{1MeVneq}. It has been obtained, assuming symmetry between the positive and negative $\mu$ beams, by reflecting the values obtained with $\mu^-$ beam around the IP and averaging direct and reflected maps.

Assuming $2\times10^{12}$ muons per bunch, 200~days of operation per year and $100$~kHz bunch crossing frequency, the 1-MeV-neq fluence results to be in the range of few $10^{15}$/cm$^2$/y in all the inner tracker and in the  electromagnetic calorimeter, steeply decreasing outside.

The total ionizing dose reaches $\sim 10^6$~Gy/y in the innermost barrel and endcap silicon layers, while it is $\sim 5\times10^4$~Gy/y in the first layers of the electromagnetic calorimeter.

These estimates have to be considered as preliminary, subject to changes in the geometry and optics, and do not include the contribution of beam--beam interactions. 

\section{Conclusion}

Among the challenges presented by a MC, a prominent role is played by the BIB as illustrated in this paper.
A massive reduction of the BIB can be obtained by means of a careful optimization of the MDI, for which a flexible simulation approach using the combination of LineBuilder and FLUKA. This software combination allows seamless linking between the code used for lattice and optics design (e.g.~MAD--X) of the MC and the Monte Carlo simulation used to assess the BIB impact on the detector.
The results for the 1.5~TeV CM energy case have been shown and compared to those obtained by the MARS15 simulations. Good agreement has been found and the residual differences are most probably due to the details of the geometry model used for some parts of the machine (i.e.~passive absorbing materials between magnetic elements). 
This paper demonstrates that this tool has the required characteristics of flexibility and accuracy in modeling the MDI as part of an integrated machine optics and MDI optimization. The next steps planned are for the study of the 3~TeV machine and its interaction region, which will also provide insights into higher energy configurations, such as a 10~TeV collider.

\acknowledgments
We acknowledge the support by the Istituto Nazionale di Fisica Nucleare and CERN. The research leading to these results has received funding from the project LUCC\_SID18\_01 of the Department of Physics and Astronomy of the University of Padova. US efforts in support of this analysis framework were supported under US Department of Energy Contracts DE-SC0012704 and DE-AC02-07CH11359.

\end{document}